\documentstyle[12pt]{article}
\begin{document}

\begin{flushright}
November 2000

OU-HET 369
\end{flushright}

\begin{center}

\vspace{5cm}
{\Large A Note on Singular Black Holes} 

\vspace{2cm}
Takao Suyama \footnote{e-mail address : suyama@funpth.phys.sci.osaka-u.ac.jp}

\vspace{1cm}

{\it Department of Physics, Graduate School of Science, Osaka University, }

{\it Toyonaka, Osaka, 560-0043, Japan}

\vspace{4cm}

{\bf Abstract} 

\end{center}

We reconsider entropy of black holes which do not have finite area horizon. 
It is suggested that some of them should have nonzero entropy from both supergravity 
and string theory point of view. 
We also refine our arguments in our previous papers for the existence of the microstates 
of the black hole.

\newpage

\vspace{1cm}

{\bf {\large 1. Introduction}}

\vspace{5mm}

One of the impressive progress in string theory is the microscopic description of black 
hole thermodynamics. 
The Bekenstein-Hawking entropy for a supersymmetric black hole is derived from string 
theory point of view, including numerical factor \cite{entropy}. 
Such black hole corresponds in string theory to a combined system of D-branes and strings, 
and therefore the entropy is the logarithm of the number of states of the brane-string 
system. 
The success of this derivation can be regarded as a quantitative test for string theory as 
the theory including gravity. 

One can consider various D-brane systems which preserve a part of supersymmetries  
\cite{solution}, but some of them correspond in supergravity to solutions which do not 
have finite area horizon. 
At first sight, this leads one to conclude that the entropy of the black hole is zero. 
However the D-brane systems can have a large number of degenerate states and the entropy of 
this system should be nonzero. 

In our previous papers \cite{previous}, we discussed D4-D4'-D0 system and argued that 
the corresponding black hole had nonzero entropy. 
We looked for classical BPS solutions in the worldvolume theory of D4-D4' intersection, 
which are regarded as D0-branes. 
The entropy can be calculated by counting the ground states of the BPS solutions and it is 
expected to be a large quantity. 
Unfortunately there is a subtlety for the existence of the BPS solutions. 

In this paper, we reconsider the black holes which have zero-area horizon and discuss that 
the entropy should be nonzero by the following reasons; (i) solutions in higher 
dimensional supergravity are regular and have nonzero radius of the horizon, (ii) stringy 
calculation provides a large number of degenerate states, (iii) improved arguments for 
D4-D4'-D0 system shows the existence of the BPS solutions. 

This paper is organized as follows. 
In section 2 we revisit a five dimensional black hole and its entropy, including the 
zero-area horizon case. 
Section 3 shows evidence for the nonzero entropy from string theory point of view. 
In section 4 we apply the same argument to a four dimensional black hole. 
We also refine the arguments for the existence of the black hole microstates.

\vspace{1cm}

{\bf {\large 2. A singular black hole in five dimensions}}

\vspace{5mm}

The most intensively studied black hole solution (or D-brane background) may be the D1-D5 
system with KK momentum. 
The metric and the dilaton field are the followings \cite{entropy},
\begin{eqnarray}
ds^2 &=& f_1^{-\frac12}f_5^{-\frac12}(-dt^2+dx_9^2+k(dt-dx_9)^2) \nonumber \\
&& \hspace{5mm} +f_1^{\frac12}f_5^{\frac12}(dx_1^2+\cdots +dx_4^2)
+f_1^{\frac12}f_5^{-\frac12}(dx_5^2+\cdots +dx_8^2) \nonumber \\
e^{-2(\phi_{10}-\phi_\infty)} &=& f_5f_1^{-1} \label{10dim}
\end{eqnarray}
where $f_1,f_5,k$ are defined as follows. 
\begin{equation}
f_1=1+\frac{c_1Q_1}{x^2}, \hspace{1cm} f_5=1+\frac{c_5Q_5}{x^2}, \hspace{1cm} 
 k=\frac{c_PN}{x^2} \hspace{1cm} x^2=x_1^2+\cdots x_4^2
\end{equation}
This D-brane system is a five dimensional object and, when they are wrapped on a five 
dimensional compact manifold, this can be regarded as 
a black hole in five dimensions. 
The corresponding solution in the five dimensional supergravity can be obtained through 
the dimensional reduction \cite{reduction}, 
\begin{equation}
ds^2_E = -\frac1{(f_1f_5(1+k))^{\frac23}}dt^2+(f_1f_5(1+k))^{\frac13}(dx_1^2+\cdots+dx_4^2)
\label{5dim}
\end{equation}
where the subscript $E$ means that this is the Einstein metric. 
This is an extremal charged black hole solution and its entropy can be calculated by the 
Bekenstein-Hawking formula \cite{BH}, 
\begin{equation}
S = \frac A{4G^{(5)}} = 2\pi\sqrt{Q_1Q_5N}. \label{entropy}
\end{equation}

If one sets $N=0$ in eq.(\ref{entropy}), the entropy of the black hole (\ref{5dim}) seems 
to vanish. 
This can be seen from the solution itself. 
Near the horizon $(x=0)$, $N=0$ solution behaves as 
\begin{eqnarray}
ds^2_E &\to& -\left(\frac xR\right)^{\frac83}dt^2+\left(\frac Rx\right)^{\frac43}
             (dx_1^2+\cdots +dx_4^2) \nonumber \\
       &=& -\left(\frac xR\right)^{\frac83}dt^2+\left(\frac Rx\right)^{\frac43}dx^2
           +R^{\frac43}x^{\frac23}d\Omega_3^2,
\end{eqnarray}
where $R$ is a constant. 
The coefficient of $d\Omega_3^2$ vanishes as $x\to 0$. 
This means that the radius of the horizon is zero, which leads us to conclude that the 
entropy of this black hole is zero. 
However, the scalar curvature of the $N=0$ solution is singular at $x=0$, and thus the 
geometric data at that region is not reliable. 
Note that the scalar curvature of the $N>0$ solution is regular even at the region. 

One may think that there is no problem; 
the $N=0$ solution has a naked singularity and should be excluded from the consideration 
by the cosmic sensorship principle. 
From string theory point of view, on the other hand, the system now considered is just 
a combination of D1-branes and D5-branes. 
Therefore there is no reason to exclude such a system from the theory. 
Moreover the D1-D5 system seems to have many degenerate bound states and this may provide 
nonzero entropy. 

\vspace{2.55mm}

Let us get back to the ten dimensional solution (\ref{10dim}). 
The near-horizon geometry for the $N=0$ solution is as follows. 
\begin{equation}
ds^2 = \frac{x^2}{\sqrt{c_1c_5Q_1Q_5}}(-dt^2+dx_9^2)+\frac{\sqrt{c_1c_5Q_1Q_5}}{x^2}dx^2
       +\sqrt{c_1c_5Q_1Q_5}d\Omega_3^2+\cdots
\end{equation}
Interestingly, this geometry is regular everywhere and has finite radius of $S^3$ 
(the coefficient of $d\Omega_3^2$). 

The D1-D5 system has their origin in ten dimensional theory, so it is natural to expect 
that the ten dimensional solution possesses correct properties of the system. 
This seems to suggest that even from the supergravity point of view the D1-D5 system with 
no KK momentum can be regarded as a five dimensional black hole with nonzero entropy.

\vspace{1cm}

{\bf {\large 3. Stringy analysis of the entropy}}

\vspace{5mm}

The low energy dynamics of the black hole given by the metric (\ref{5dim}) can be described 
by a two dimensional superconformal field theory living on the D1-branes \cite{entropy}.  
If the five dimensional space is compactified to $M_4\times S^1$, where the D1-branes are 
wrapped on $S^1$, the target space $X$ of the two dimensional SCFT is a symmetric product 
of $M_4$ i.e. $X=S^kM_4$. 
The entropy of the black hole can be calculated by counting BPS states with appropriate 
weight in the SCFT and taking the logarithm of the degeneracy of the states. 
A useful tool to do the task is the elliptic genus of $X$ \cite{genus},
\begin{equation}
\chi(X;q,y) = Tr_{RR}(-1)^Fy^{F_L}q^{L_0-\frac d8}
\end{equation}
where $F=F_L+F_R$ is the total fermion number and $d=\mbox{dim}_{\bf C}X$. 
This is the weighted sum of the number of states whose right-moving states are restricted 
to the ground states, which enables us to take into account only BPS states. 
The trace is taken only over RR-sector because the black hole geometry admits the periodic 
Killing spinors \cite{Killing} and this indicates that the corresponding states are in 
RR-sector. 
Surprisingly, such calculation reproduces the correct entropy formula (\ref{entropy}), 
including the numerical factor. 

\vspace{2.5mm}

Now consider the $N=0$ black hole. 
For the same reason as in the $N>0$ case, this black hole should correspond to states in 
RR-sector. 
Moreover, these states are the ground states of RR-sector \cite{Killing}. 
The contribution from the ground states can be picked up by setting $q=0$. 
The result is \cite{genus}
\begin{equation}
\chi(X;q=0,y) = \sum_{r,s}(-1)^{r+s}h^{r,s}(X)y^{r-\frac d2}.
\end{equation}
Thus the number of the ground states is equal to the sum of the Hodge numbers of the target 
space manifold $X$. 
For example, if $M_4=K3$ then $X=S^{Q_1Q_5+1}K3$ and 
\begin{equation}
\sum_{r,s}h^{r,s}(X) = \sum_kb_k(X) \sim e^{4\pi\sqrt{Q_1Q_5+1}},
\end{equation}
for large $Q_1Q_5$. 
Therefore the entropy should be 
\begin{equation}
S=4\pi\sqrt{Q_1Q_5+1}.
\end{equation} 
This ground state degeneracy is also discussed in \cite{degeneracy}. 
Note that this contribution to the entropy is negligible compared with eq.(\ref{entropy}) 
when $N$ is very large, which is the limit necessary for the derivation of 
eq.(\ref{entropy}) from SCFT.

\vspace{1cm}

{\bf {\large 4. A singular black hole in four dimensions}}

\vspace{5mm}

In this section, we apply the similar argument to a four dimensional black hole. 
In \cite{previous}, we considered a black hole which corresponds to D4-D4'-D0 brane system. 
The supergravity solution of the brane system has zero-area horizon, but there exists the 
curvature singularity like the five dimensional case. 
Therefore it may be appropriate for this case to be discussed in a higher dimensional theory. 

After taking T-duality and lifting to M-theory, the brane system corresponds to an 
intersecting M5-brane system. 
Its supergravity solution is as follows \cite{solution}. 
\begin{eqnarray}
ds_{11}^2 &=& (H_1H_2H_3)^{\frac23}\Bigl\{(H_1H_2H_3)^{-1}(-dt^2+dx_{10}^2)+dx_1^2+dx_2^2
               +dx_3^2 \nonumber \\
          & &  \hspace{-5mm}+(H_1H_3)^{-1}(dx_4^2+dx_5^2)+(H_1H_2)^{-1}(dx_6^2+dx_7^2)
               +(H_2H_3)^{-1}(dx_8^2+dx_9^2)\Bigr\} 
\end{eqnarray}
$$ H_k = 1+\frac{c_kQ_k}r \hspace{5mm} (k=1,2,3), \hspace{1cm} r^2 = x_1^2+x_2^2+x_3^2.  $$
Its geometry near $r=0$ is 
\begin{equation}
ds_{11}^2 \to \left(\frac\rho{\rho_0}\right)^2(-dt^2+dx_{10}^2)+
              \left(\frac{\rho_0}\rho\right)^2d\rho^2
              +\left(\frac{\rho_0}2\right)^2d\Omega_2^2+\cdots,
\end{equation}
where $\rho=\sqrt{2\rho_0r}$ and $\rho_0$ is a constant. 
This is regular everywhere and has nonzero radius of $S^2$, and thus this black hole 
is expected to have nonzero entropy. 

One can also find a regular solution in Type IIA supergravity,
\begin{eqnarray}
ds_{10}^2 &=& f_4^{-\frac12}f_{4'}^{-\frac12}(-dt^2+dx_9^2)+f_4^{\frac12}f_{4'}^{\frac12}f_5
              (dx_1^2+dx_2^2+dx_3^2)+f_4^{-\frac12}f_{4'}^{-\frac12}f_5dx_4^2 \nonumber \\
          & & +f_4^{-\frac12}f_{4'}^{\frac12}(dx_5^2+dx_6^2)+f_4^{\frac12}f_{4'}^{-\frac12}
              (dx_7^2+dx_8^2), 
\end{eqnarray}
$$ f_4=1+\frac{c_4Q_4}r,\hspace{1cm} f_{4'}=1+\frac{c_{4'}Q_{4'}}r,\hspace{1cm} 
f_5=1+\frac{c_5Q_5}r, $$
which corresponds to D4-D4'-NS5 system and is related to D4-D4'-D0 system by U-duality. 
The entropy may be obtained by using CFT technique as in the previous section. 

In our previous papers \cite{previous}, we investigated the spacetime effective theory of 
D4-D4' system living on the D4-D4' intersection and tried to find out BPS states of the 
theory, which correspond to the D0-branes, as classical monopole solutions. 
However, the BPS equation does not have regular solution. 
The difficulty to find the BPS solutions will be related to the fact that the system 
we considered is a marginal bound state, and 
therefore the quantum effects may be important for this situation. 

A way to discuss the existence of the states is to deform the theory by which the marginal 
bound states become truly bound states \cite{deform}. 
For definiteness we consider the D4-D4' system in which a D4-brane is extended along (04568) 
directions and a D4'-brane is along (04579) directions. 
This is a supersymmetric configuration and has no tachyonic mode which, if exists, 
indicates the instability of the system. 
If the D4'-brane is rotated in (89)-plane by angle $\phi$, it is well-known that the 
cancellation of zero-point energy does not occur and there emerges a tachyonic mode. 
The $\mbox{mass}^2$ of the tachyon is 
\begin{equation}
m^2 = -\frac{\nu}{2\alpha'}, \hspace{1cm} \nu = \frac{\phi}{\pi}.
\end{equation}
Now we would like to take $\alpha'\to 0$ limit to decouple the bulk gravity and the higher 
excited modes from the worldvolume theory while keeping the tachyon $\mbox{mass}^2$ finite. 
Such limit can be taken as 
\begin{equation}
\alpha'\to 0, \hspace{5mm} \phi\to 0, \hspace{5mm} \frac\phi{\alpha'}=\mbox{fix}=2\lambda.
\end{equation}
After taking this limit, there remain the following fields in addition to ${\cal N}=4$ 
vectormultiplet in three dimensions.

\vspace{2mm}

\hspace*{1cm}$\cdot$ a complex tachyonic scalar with $m^2=-\lambda$

\hspace*{1cm}$\cdot$ a complex massive scalar with $m^2=+\lambda$

\hspace*{1cm}$\cdot$ massless fermions (same degrees of freedom with the above scalars)

\vspace{2mm}

Since we take $\phi\to 0$, this system should recover eight supercharges. 
The effective theory of this system with the above fields is three dimensional ${\cal N}=4$ 
gauge theory coupled to a hypermultiplet with the FI parameter $\lambda$. 
This theory has the vacuum in the Higgs branch and the vortex which preserves half of 
supersymmetries. 
By the nonrenormalization theorem in the Higgs branch \cite{nonrenorm}, this vortex exists 
also in the quantum level, and this will indicate the existence of the D4-D4'-D0 bound 
states. 

Consider now the combined system of $Q_1$ D4-branes, $Q_2$ D4'-branes and $N$ D0-branes. 
Naively, the entropy of the corresponding black hole can be calculated by counting the 
number of ways of distribution of $N$ D0-branes into $Q_1Q_2$ D4-D4' intersections. 
This number grows rapidly as $N$ becomes large. 
The logarithm of this is proportional to $\sqrt{Q_1Q_2N}$. 
This shows that there are a large number of degenerate states which will correspond to 
the black hole microstates. 
The determination of the entropy formula, from this deformed theory, including the 
proportionality factor will need to study the properties of the vortices in non-Abelian 
gauge theory. 

\vspace{2cm}

{\bf {\large Acknowledgments}}

\vspace{5mm}

I would like to thank R-G.Cai, H.Itoyama, T.Matsuo, K.Murakami, N.Ohta, Y.Okawa for 
valuable discussions. This work is supported in part by JSPS Reseach Fellowships.

\end{document}